\documentclass[showpacs,aps,prl,twocolumn,superscriptaddress,10pt]{revtex4}
\usepackage{graphicx} 
\usepackage{bm}
\usepackage{color}

\usepackage{amsfonts}
\usepackage{amsmath}
\usepackage{amssymb}
\usepackage{enumerate}
\usepackage{xspace}
\usepackage{mathrsfs}
\usepackage{mathptmx}
\usepackage[colorlinks=true,linkcolor=blue,citecolor=blue,urlcolor=blue]{hyperref}
\usepackage[utf8]{inputenc}

\usepackage[T1]{fontenc}
\usepackage{graphicx}

\usepackage{textcomp}
\usepackage{amsmath,amsfonts}
\usepackage{lineno}
\usepackage{amssymb}
\usepackage{float}
\usepackage{amsfonts}
\usepackage{amssymb}
\usepackage{amsmath}
\usepackage{physics}
\usepackage{upgreek}
\usepackage{textcomp}
\usepackage{xcolor}

\usepackage{placeins}

\usepackage{color,soul}

\newcommand*\diff{\mathop{}\!\mathrm{d}}

\begin{document}

\title{Topological energy release from collision of relativistic  antiferromagnetic solitons}
\author{R. M. Otxoa}
\email{ro274@cam.ac.uk}
\affiliation{Hitachi Cambridge Laboratory, J. J. Thomson Avenue, Cambridge CB3 0HE, United Kingdom}
\affiliation{Donostia International Physics Center, 20018 San Sebasti\'an, Spain}
\author{R. Rama-Eiroa}
\affiliation{Donostia International Physics Center, 20018 San Sebasti\'an, Spain}
\affiliation{Polymers and Advanced Materials Department: Physics, Chemistry, and Technology, University of the Basque Country, UPV/EHU, 20018 San Sebasti\'an, Spain}
\author{P. E. Roy}
\affiliation{Hitachi Cambridge Laboratory, J. J. Thomson Avenue, Cambridge CB3 0HE, United Kingdom}
\author{G. Tatara}
\affiliation{RIKEN  Center  for  Emergent  Matter  Science  (CEMS)  and  RIKEN  Cluster  for Pioneering  Research  (CPR),  2-1  Hirosawa,  Wako,  Saitama,  351-0198  Japan}
\author{O. Chubykalo-Fesenko}
\affiliation{Instituto de Ciencia de Materiales de Madrid, CSIC, Cantoblanco, 28049 Madrid, Spain}
\author{U. Atxitia}
\affiliation{Dahlem Center for Complex Quantum Systems and Fachbereich Physik, Freie Universitat Berlin, 14195 Berlin, Germany}

\begin{abstract}
Magnetic solitons offer functionalities as information carriers in multiple spintronic and magnonic applications. However, their potential for nanoscale energy transport has not been revealed. Here we demonstrate that antiferromagnetic solitons, e.g. domain walls, can uptake, transport and release energy. The key for this functionality resides in their relativistic kinematics; their self-energy increases with velocity due to Lorentz contraction of the soliton and their dynamics can be accelerated up to the effective \textit{speed of light} of the magnetic medium. Furthermore, their  classification in robust topological classes allows to selectively release this energy back into the medium by colliding solitons with opposite topology. Our work uncovers important energy-related aspects of the physics of antiferromagnetic solitons and opens up the attractive possibility for spin-based nanoscale and ultra-fast energy transport devices.

\end{abstract}

\maketitle


Solutions for an efficient control of energy in nanoelectronics are based on identifying the prevailing carriers and transfer mechanisms of energy at relevant time and length scales. 
The possibility of using spin degrees of freedom as energy carriers is barely known.
Localized magnetic textures --  non-collinear spin structures -- such as domain walls  (DWs), vortices and skyrmions, whose stability is grounded to their non-trivial topology \cite{bogdanov2001chiral, roessler2006spontaneous,LitziusNatElectronics2020}, have  been already discussed broadly as information carriers \cite{parkin2008magnetic,allwood2002submicrometer} but rarely as stationary energy storing devices \cite{An2013NatureMat,VedmedenkoPRL2014,TserkovnyakPRL2018,DaltonPRB2020}. 
 These so-called magnetic solitons (MSs) already play a pivotal role for the development of spin-based applications such as processing \cite{parkin2008magnetic,allwood2002submicrometer}, sensing \cite{allwood2005magnetic}, storing information \cite{fukami2009low} as well as radio-frequency \cite{pribiag2007magnetic} and neuro-inspired devices \cite{7563364,grollier2020neuromorphic}. The simplest example of a MS is the DW, which separates magnetic domains magnetized in opposite directions \cite{hubert2008magnetic}. The exchange energy stored in the DW can be transported by moving the DW with magnetic fields \cite{schryer1974motion}, spin currents \cite{slonczewski1996current,miron2011fast}, spin waves \cite{han2009magnetic,wang2012domain,yan2011all} or even thermal gradients \cite{torrejon2012unidirectional,PhysRevLett.117.107201}. While their strong stability allows for long-range information and energy transport, it impedes releasing their stored topological energy. Moreover, the only way to modify the energy of a DW on demand is by controlling its width. 
 However, a DW has a self-energy that is defined by intrinsic material dependent magnetic parameters, such as exchange and anisotropy energy terms. Thus, means for modifying, transporting and extracting the free energy at magnetic textures are missing.

    
\begin{figure}[!ht]
\centering
\includegraphics[width=8.6cm]{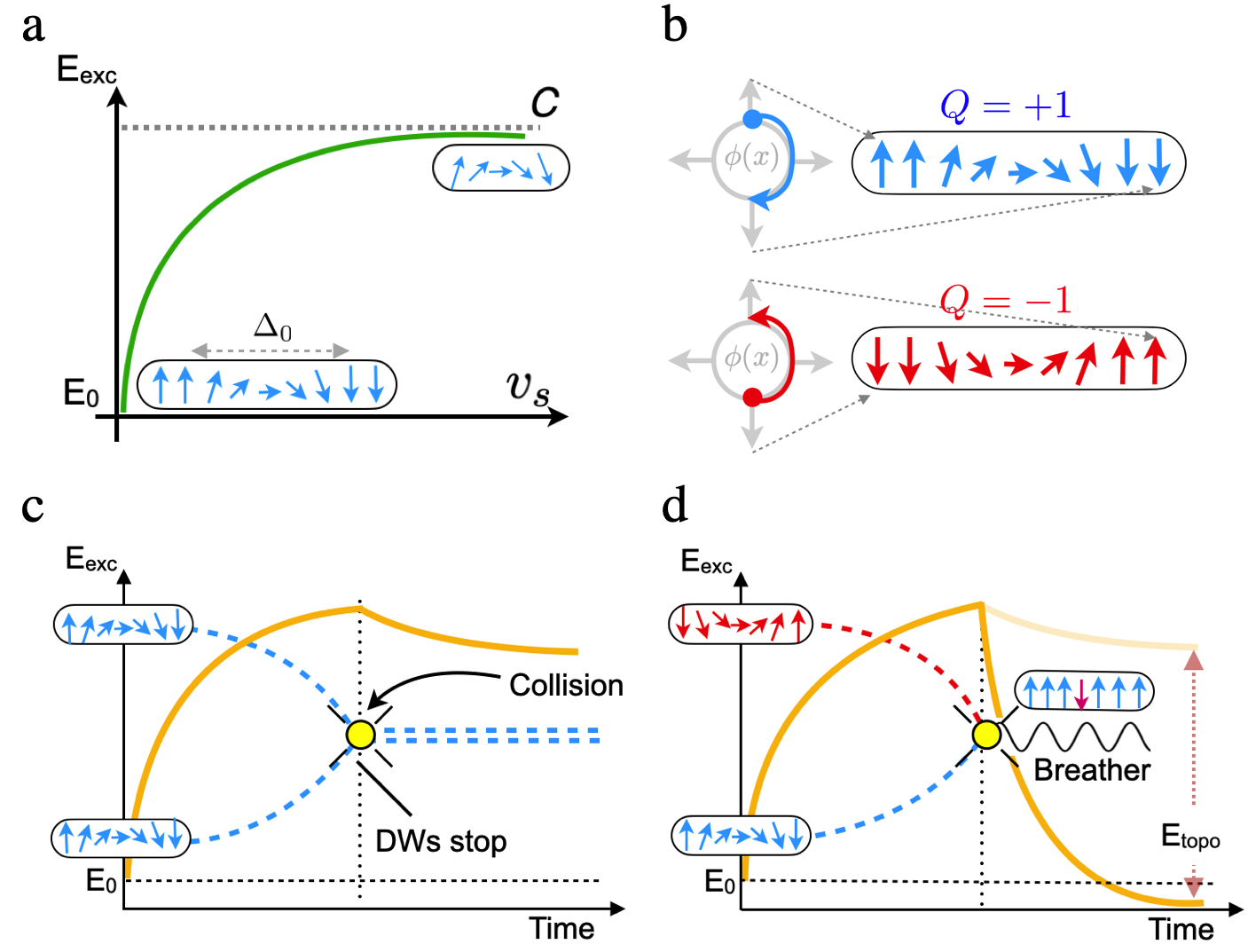}
\caption{
(a) In antiferromagnets, domain walls (DW) can be driven up to relativistic speeds $v_s \approx c$, resulting in the Lorentz-contraction of the DW, $\Delta=\Delta_0\gamma_{\rm{L}}$, and the increase of the self-energy, $E(v_s)=E_0/\gamma_{\rm{L}}$, where $\gamma_{\rm{L}}=\sqrt{1-(v_s/c)^2}$.  
(b) DWs can hold two spin winding numbers $Q= \pm 1$, which are degenerated in energy. 
(c) two DWs with $Q=+1$ and $Q=+1$, approaching each other in an accelerated motion until collision. After collision DWs stop. 
(d) two DWs with the opposite winding number, $Q=+1$ and $Q=-1$, approaching each other in an accelerated motion until collision. After collision DWs annihilate, a breather appears, and topological energy $E_{\rm{topo}}$ released.
}
\label{im:0}
\end{figure}

 Our proposal relies on the unique dynamical properties of DWs (MSs) in antiferromagnetic materials (AFM) \cite{Gomonay2014Review,BaltzRevModPhys2018,Gomonay2018NatPhys,PhysRevLett.117.087203,OtxoaCommPhys2020}. Since they obey the relativistic kinematics, their width and energy strongly depend on the soliton velocity (Fig. \ref{im:0}a), which results in a significant increase of magnetic energy in the system. Differently to DWs in ferromagnets which are prone to deformation at relatively low velocities \cite{hubert2008magnetic,Mougin2007EPL,hertel2006exchange, PhysRevB.101.224425}, AFM DWs offer the possibility to transport their self-energy at speeds close to the effective \textit{speed of light} of the medium, $c$ \cite{PhysRevLett.50.1153} (Fig. \ref{im:0}a). As they preserve their shape through time, these  MSs enable long-range coherent energy transport \cite{Mohseni2013Science,rajaraman1982solitons}. Notably, since the relative orientation of atomic spins within the DWs can be classified into two distinct topological classes (Fig. \ref{im:0}b), here we demonstrate the possibility of topologically-mediated energy release by collision of two relativistic AFM DWs (Figs. \ref{im:0}c and d). For instance, for a one dimensional (1D) DW, the topological charge is defined as the integral over the space of the winding number density, $w\left(x,t\right)=-\nabla_{x}\phi\left(x,t\right)$. Here $\phi\left(x,t\right)$ is the in-plane angle of the spin at location $x$  at time $t$ of the spin-configuration along the 1D-line of the propagation direction along the $x$-axis (Fig. \ref{im:0}b). The total winding number or topological charge is expressed as: $Q= \frac{1}{\pi}\int w\left(x,t\right)dx$. Therefore, a DW in a system can exist with only two distinct topological flavours, $Q=\pm 1$ (Fig. \ref{im:0}b). It is important to note that the homogeneous state corresponds to $Q_0=0$, since $\phi\left(x,t\right)=$ constant.
When two DWs of the same topology ($Q_1=+1$ and $Q_2=+1$) are driven to collide, they cannot collapse into an homogeneous state (Fig. \ref{im:0}c). 
However, when at least two DWs with opposite topological charge ($Q_1=+1$ and $Q_2=-1$) are forced to collide (Fig. \ref{im:0}d) a recombination process is accessible both from energy and topological arguments as now $Q_0=Q_1+Q_2=0$. Importantly, the topologically protected energy stored at the DWs, $E_{\rm{topo}}$ is completely released.

In AFMs, the role played by the photons in special relativity corresponds to the magnons, and  $c$ corresponds to their maximum group velocity. A direct consequence of \textit{special relativity} is that the MSs' width at rest, $\Delta_0$, contracts as its velocity, $v_{\text{s}} \rightarrow c$, described by Lorentz  factor, $\gamma_{\rm{L}}$;   $\Delta=\gamma_{\rm{L}} \Delta_0$, where $\gamma_{\rm{L}}(v_{\text{s}})=\sqrt{1-(v_{\text{s}}/c)^{2}}$. An immediate consequence is that MSs' "kinetic" energy can also be represented in a relativistic form given by 
\begin{equation}
    E_{\text{exc}}(v_{\text{s}})=\frac{E_0}{\sqrt{1-(v_{\text{s}}/c)^{2}}},
    \label{Eq:1}
\end{equation}

\noindent where $E_0$ corresponds to the MSs' energy at rest (Fig. \ref{im:0}a).
Importantly, relativistic soliton physics is not only a theoretical construct but indirect experimental verification has been recently achieved in ferrimagnetic insulators \cite{Caretta2020Science}. 
Although propagation at relativistic velocities of individual AFM DWs have been investigated theoretically, their interactions with each other and the role of topology remain unexplored. 
 \begin{figure}[!t]
\includegraphics[width=8.6cm]{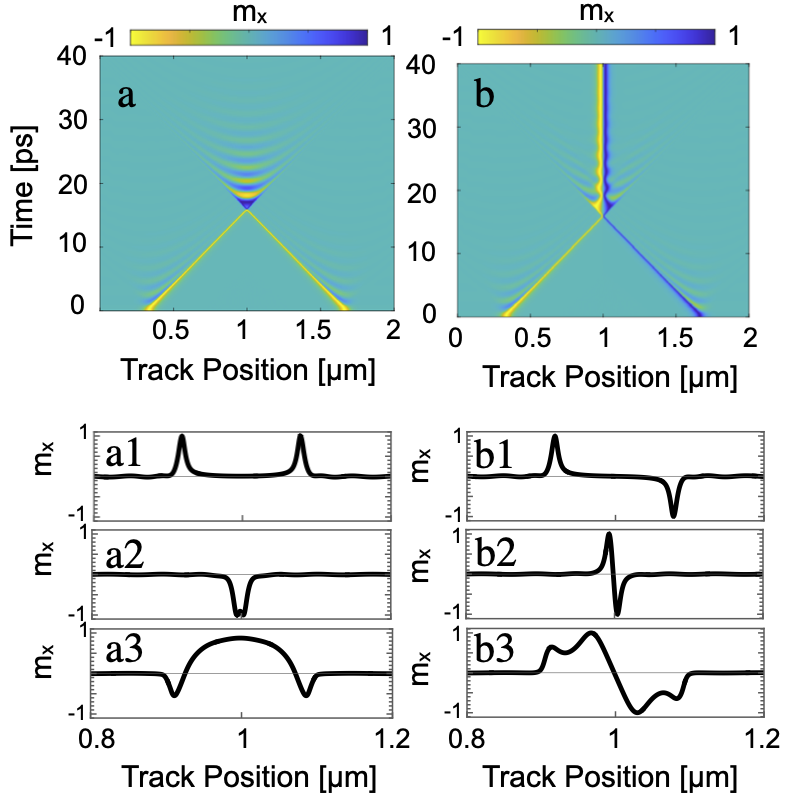}
\caption{Model simulations of the collision of two domain walls (DWs) in Mn$_2$Au. 
 (a,b) Spatio-temporal diagram of the dynamics of the process showing the DWs collision and the corresponding output. (a) Two DWs with opposite topological charges result after the collision in a dispersing breather; (b) Two DWs with the same topological charges after the collision stop and increase their size. A finite $x$-component of the magnetization, $m_x$, represents the position and extension of the DWs. Subplot 
(a1,b1), (a2,b2) and (a3,b3) are  cuts of the colour map in (a) and (b) for three characteristic times $t=14,16$ and 18 ps, respectively. 
} 
\label{im:2}
\end{figure}


We quantify and determine the energy flow dynamics associated to the DW motion and the collision of two high energy DWs in the AFM metal Mn$_2$Au. This material presents a high N\'{e}el temperature of circa 1500 K \cite{barthem2013revealing}, and an efficient electric control of the DW motion  \cite{Manchon2019RevModPhys,OtxoaCommPhys2020,Otxoa2020CommPhys}. Upon passing an electrical current along the basal planes, the so-called inverse spin galvanic effect \cite{ganichev2008spin} produces a staggered local spin accumulation with opposite polarities in each sublattice which creates a local staggered spin-orbit (SO) field perpendicular to the  current direction. The dynamics of a DW induced by SO-torques can be described by the Landau-Lifshitz-Gilbert (LLG) equation for the atomistic spin dynamics (described in online Supplemental Material). To follow the motion of a DW in a stripe with a long dimension parallel to $x$-axis,
 it is sufficient to monitor the $m_x$ projection of the magnetization along the track. The initial condition in our computational model corresponds to two DWs separated by approximately 1.5 $\mu$m. The SO-field generated by the laterally injected electrical current acts onto the DWs through the Zeeman energy. In order to reduce the Zeeman energy, the magnetic domain between the two DWs shrinks, moving both DWs towards each other. The DWs reach their final velocities of around 42.4 km/s for a SO-field of 60 mT with a ramping time of 10 fs in only a couple of picoseconds. This speed constitutes $\sim$98$\%$ of the maximum magnon velocity, $c$, for Mn$_2$Au. For the range of SO-fields investigated, the DW width reduces from 20 nm to 4 nm due to the Lorentz contraction. This leads to an increase of a 500\% of the DWs energy which can be transported (Fig. \ref{im:0}a and Supplementary Material). Notably, DWs are very stable with no visible spinwave generation at these high velocities. Circa 15 ps later both DWs reach each other (Fig. \ref{im:2}a and b) and collide. 
 Two scenarios arise depending upon the topological charge carried by them, and topology conservation laws.


 \begin{figure}[!t]
\includegraphics[width=8.9cm]{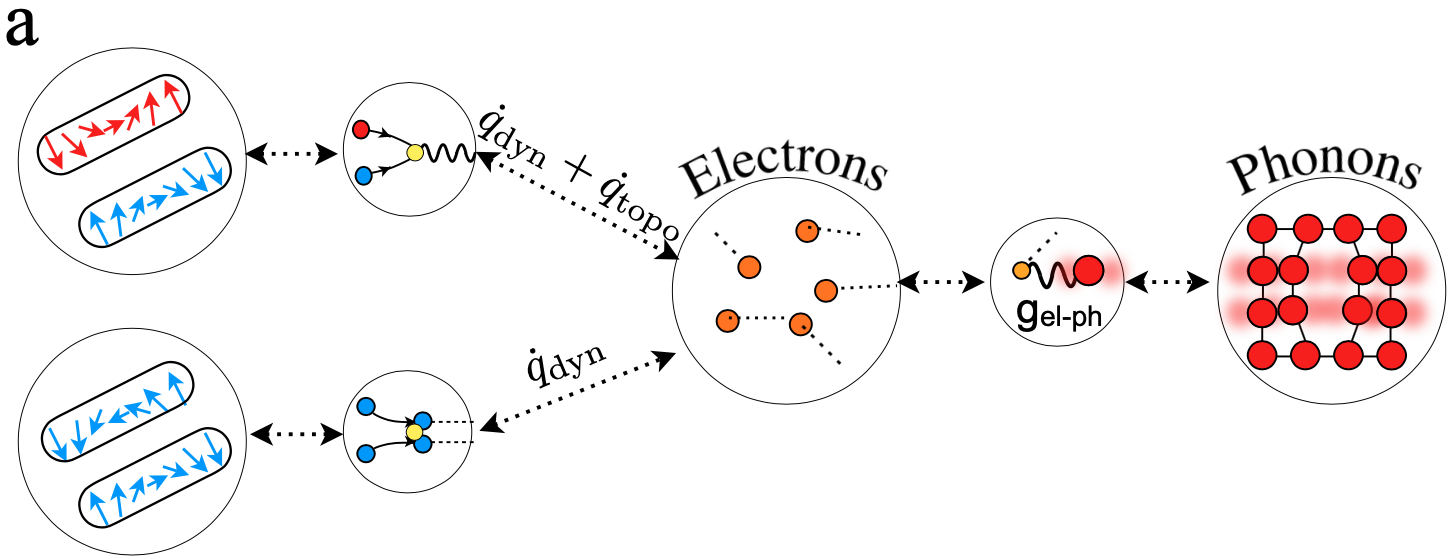}
\includegraphics[width=8.6cm]{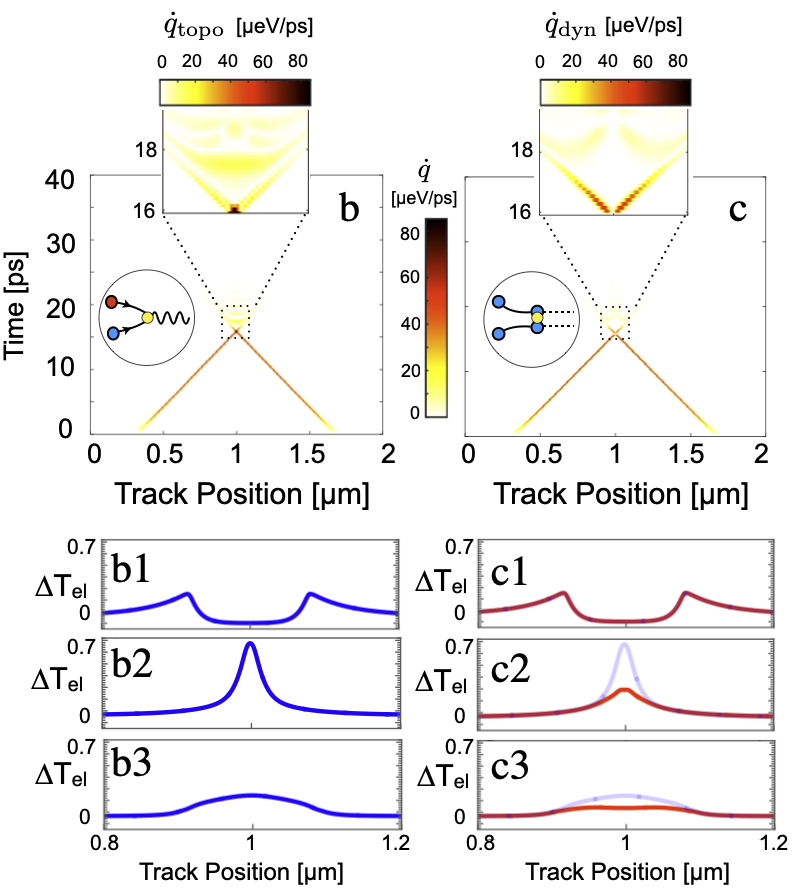}
\caption{(a) Stored-energy at two domain walls (DWs) with opposite topology can be transferred to the electron system by colliding them. Besides $\dot{q}_{\rm{dyn}}$, an additional topologically-mediated transfer mechanisms opens up, $\dot{q}_{\rm{topo}}$, when the winding of the DWs are opposite.  
The electron and lattice are coupled via electron-phonon coupling, $g_{\rm{el-ph}}$. (b) Colour map of the spatio-temporal distribution of the rate of heat dissipation caused by the motion of the DWs. At the collision (zoomed in inset) an explosion of heat occurs due to the recombination of DWs. Subplots (b1/c1), (b2/c2) and (b3/c3) represent the spatial distribution of $\Delta T_{\rm{el}}$ along the track for three characteristic times: 14, 16 and 18 picoseconds, respectively.} 
\label{im:3}
\end{figure}

Specifically, when $Q_1=-Q_2=1$, topological charge conservation rules allow  the DWs to recombine leading to a homogeneous state $Q_0=0$ (Fig. \ref{im:2}a). At an instant prior to the collision ($t=14$ ps),  both DWs are well defined (Fig. \ref{im:2}a1), close to the moment at which they annihilate each other ($ t = 16 $ ps)  the two DW profiles merge (Fig. \ref{im:2}a2), and some time after their disappearance as individual entities ($t=18$ ps),  a bounded dispersing stationary breather mode is observed (Fig. \ref{im:2}a3). 

Interestingly, a breather mode \cite{rajaraman1982solitons} is created out of the collision between the two solitons, which is localized in the space-time by a cone defined by the trajectory of each DW before the collision  (Fig. \ref{im:2}a). No spin perturbations can exist outside this cone. We observe that the past magnon-cone behaves like the future magnon-cone in reverse, however, differently to relativity, spin perturbations lie behind the DW motion, namely outside the cone \cite{PhysRevResearch.2.043226}. Importantly, the breather mode attenuation time  is given by the exchange relaxation time scale: $2\alpha_{\rm{G}} \gamma J_{\rm{AFM}}/\mu_{\rm{s}} \approx 3-4$ ps, where $\alpha_{\rm{G}}$ is a phenomenological parameter, named Gilbert damping, which controls the dissipation of angular momentum in magnetic materials, $\gamma$  is the gyromagnetic ratio and $\mu_{\rm{s}}$ is the sublattice atomic magnetic moment (see SM). In layered AFM, such as Mn$_2$Au, $J_{\rm{AFM}}$ represents the effective exchange interaction between layers. This sets an ultra-fast discharge timescale. For more details of the dynamics see also Supplementary videos. 

When $Q_1=Q_2=1$, topological charge conservation rules do not allow for DWs to recombine leading to the homogeneous state ($Q_0=0$). Similar to the previous scenario, at an instant prior to the collision ($t=14$ ps), both DWs are well defined (Fig. \ref{im:2}b1). After the collision, the DWs stop their rectilinear motion (conservation of momentum) ($ t = 16 $ ps) (Fig. \ref{im:2}b2) hence the Lorentz factor $\gamma_{\text{L}}\approx 1$ and as a consequence the DWs widen almost instantaneously, but start to oscillate around their final position emitting spin waves ($t=18$ ps) (Fig. \ref{im:2}b3).
The final state, i.e. DW widths and separation, however depends on energetic considerations. The system tries to minimize the Zeeman energy at the domain in between DWs by approaching them. At the same time, the repulsive exchange interaction becomes significant when the DWs approach sufficiently. As a result of the competition between the two energies, the DWs stabilise at a certain distance between each other (30-90 nm, depending on the SO-field magnitude, i.e. the separation between DWs is much larger than their widths), see detailed calculations in the Supplementary Materials. The spin wave emission is also observed, especially during the DWs stabilisation time. 

\begin{figure}[t]
\includegraphics[width=8.5cm]{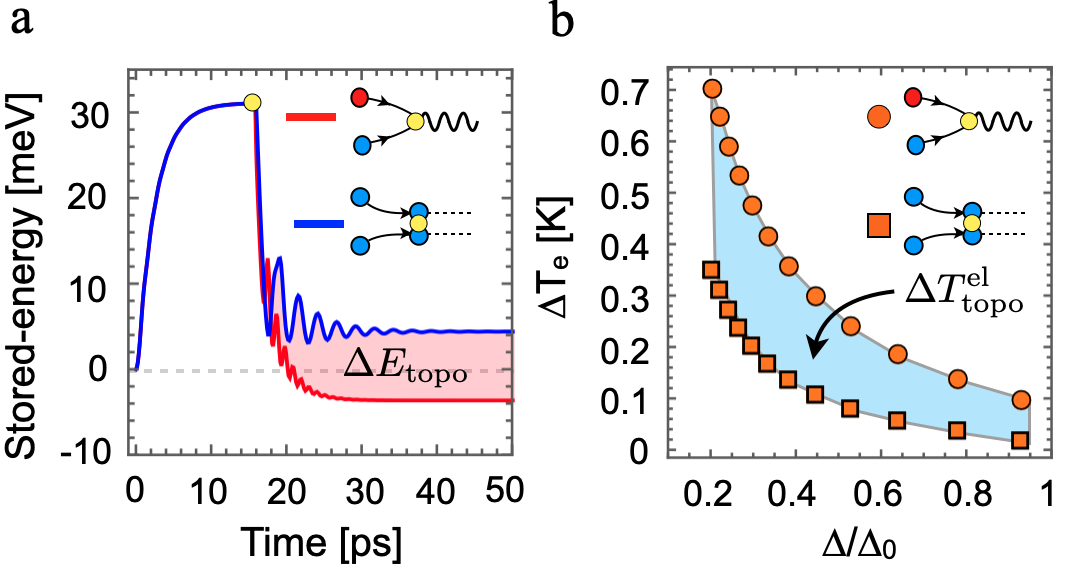}
\caption{(a) Time evolution of the stored-energy at the DWs. Before domain wall (DW) collision (yellow circle) energy increase as the velocity of the DWs accelerates. After collision (yellow circle), for $Q_1=-Q_2$, a complete release of stored-energy, while for $Q_1=Q_2$, partial release of stored-energy take place. The difference of stored-energy release is defined as $\Delta E_{\rm{topo}}=\Delta E (Q_1,-Q_1)-\Delta E (Q_1,Q_1)$.  (b) The maximum (at the collision)  temperature increase of the electron system  as a function of the reduced DW width, $\Delta /\Delta_0$. The shaded area defines the topological effect $\Delta T^{\rm{el}}_{\rm{topo}}=\Delta T_{\rm{el}} (Q_1,-Q_1)-\Delta T_{\rm{el}}  (Q_1,Q_1)$.
} 
\label{im:4}
\end{figure}

Still a fundamental question needs answer: how much energy is accessible to the external environment? 
The magnetic energy flow during the DWs collision involves the electron and lattice systems. To further quantify this process, we use a  kinetic model where both local and non-local electron, phonon and spin relaxations are included \cite{Otxoa2020CommPhys}.  The dynamics of the electron and lattice vibrations (phonons) energies (expressed as their ``effective quasi-equilibrium temperatures" which can be a measurable quantity) are described by the two temperature model (TTM) \cite{Kaganov1957} (Fig. \ref{im:4}a).  

The fundamental assumption here is that the magnetic free energy flows directly into the electron system due to metallic nature of Mn$_2$Au (Fig. \ref{im:3}a).  
\begin{eqnarray}
\label{eq:2TM-main}
C_{\rm{el}} \frac{d T_{\rm{el}}}{d t} &=& -g_{\rm{el-ph}}\left( T_{\rm{el}} - T_{\rm{ph}} \right)  + \frac{\partial}{\partial x} \kappa \frac{\partial T_{\rm{el}} }{\partial x}+\dot{q} \nonumber \\
C_{\rm{ph}} \frac{d T_{\rm{ph}}}{d t} &=& g_{\rm{el-ph}}\left( T_{\rm{el}} - T_{\rm{ph}} \right).
\end{eqnarray}
The electron system receives an input of energy from  the DW motion, due to the magnetic friction (spin-Peltier effect \cite{Otxoa2020CommPhys}), spin wave attenuation and DW collision, all of them quantified by the Rayleigh dissipation functional, $\dot{q}= \eta \dot{s}^2$, where $\eta=\mu_{\rm{s}}\alpha_{\rm{G}}/\gamma$.
 Since the electron system have a much lower specific heat ($C_{\rm{el}}= \gamma_{\rm{el}} T_{\rm{el}}$) than the lattice ($C_{\rm{ph}})$, the  electron system heats up almost instantaneously at a temperature that is larger than the lattice. The lattice heats only indirectly due to the coupling to the hot electrons via the electron-phonon coupling, $g_{\rm{el-ph}}$ (Fig. \ref{im:3}a). The lateral thermal electron diffusion (defined by the parameter $\kappa$) is also included. We assume room temperature parameters \cite{Otxoa2020CommPhys}. The details of the electron and phonon temperature dynamics can be visualised in Supplementary videos for the two topological classes presented above.

For a stationary moving DW, the maximum heat is released at the DW center and the rate of its density $\dot{q}$  is defined by  
\begin{equation}
    \dot{q}_{\rm{dyn, stat}} =\frac{\mu_{\rm{s}}}{\gamma} \alpha_{\rm{G}} \left( \frac{v_s}{\Delta}\right)^2.
\label{eq:dissipation-analytical-max}
\end{equation}
We observe  (Fig. \ref{im:3}b and c) that while the DWs move towards each other, their dissipation rates are equal, $\dot{q}_1=\dot{q}_2$, and  closely follows Eq. (\ref{eq:dissipation-analytical-max}).  Before the collision (Figs. \ref{im:3}b,c, b1 and c1), two well-defined electron temperature  peaks can be clearly observed, corresponding to the DW positions.  As DWs approach each other closer, their shapes are affected by their mutual interaction and individual topology. Spin wave generation is also observed. Remarkably, DWs positions are clearly defined until the collision, marked  by two separate $\Delta T_{\rm{el}}$ peaks.

When $Q_1=Q_2$,  at the instant of  DWs collision, due to the repulsion effect, their positions never merge (visible as a white spot in the dissipated energy in the inset of Fig. \ref{im:3}c). After the collision, DWs velocities change and the energy excess  redistributes between different subsystems, via spin wave emission, electron-phonon coupling and lateral thermal conduction. This leads  to a small energy peak (Fig. \ref{im:3}c2). Since DWs remain well-defined separate entities during the entire dynamics, we identify this process as a purely dynamical energy release, $\dot{q}_{\rm{dyn}}$,  the same as the spin-Peltier effect in  moving magnetic textures \cite{Otxoa2020CommPhys} but for a more complex situation.

On the contrary, when $Q_1=-Q_2$,  at the moment of DW relativistic collision ($t=16$ ps, Fig. \ref{im:3}b2), there is a large temperature boost (visible as a dark spot in the dissipated energy in the inset of Fig. \ref{im:3}b) which adds to the  temperature increase from magnetic friction, $\dot{q}_{\rm{dyn}}$.  Since this effect is caused by the  annihilation of TMS subject to special topological rules, we call this process topological energy release  ($\dot{q}_{\rm{topo}}$). Soon after ($t=18 $ ps, Fig. \ref{im:3}b3), we again observe attenuation of $\Delta T_{\rm{el}}$ at the collision site due to the energy redistribution, via electron-phonon coupling and lateral thermal conduction. On longer timescale the breather disperses causing lateral spin waves which  finally attenuate, passing the energy to electron and phonon systems. 

The stored-energy in the DW increases as the DW accelerates to reach their maximum velocity given by the SO-field (Fig. \ref{im:4}a).
Interestingly, the dynamics of DW energies resembles that of the electric capacitor charging and discharging. Control of the amount of energy released into the medium is  possible using high-energy AFM DWs. 
 Since the self-energy of the individual DWs follows the special relativity Eq. (\ref{Eq:1}), one can adjust it by modifying the width of the DW at the moment of the collision, $\Delta_{\mathrm{c}}$ (Fig. \ref{im:3}b). 
 The variation of the self-energy of the DWs before and after the collision, $\Delta E_{\rm{col}}$, depends on both $\Delta_{\text{c}}/\Delta_0$ and their relative winding numbers $Q_1,Q_2$. From our calculations, for relatively large velocities, close to $c$, the reduction of $\Delta$ can reach relative values of $\Delta_{\text{c}}/ \Delta_0 \approx 0.2$.  For the system parameters considered here, for  collisions events with $Q_1 Q_2<1$ occurring at the highest velocities (corresponding to a SO-field = 60 mT), we estimate the transferred energy $\Delta E_{\rm{topo}} \sim 35$ meV.  
In particular, we define topological energy variation in terms of the corresponding temperatures as 
$\Delta T^{\rm{el}}_{\rm{topo}}=\Delta T_{\rm{el}}(Q_1,-Q_1) - \Delta T_{\rm{el}}(Q_1,Q_1)$,
represented in Fig. \ref{im:4}b as a function of the DW width. We predict a measurable electron peak temperature difference between the two topologically distinct DW collision processes, $\Delta T^{\rm{el}}_{\rm{topo}}=0.35$ K. 

To summarize, we have shown that topological magnetic solitons (TMSs) can be used as energy carriers. In particular, antiferromagnetic domain walls (AFM DWs) can uptake, transport and release energy. These mechanisms are universal for any type of AFM TMS (skyrmions, vortices, etc.). 
In a first step, we have demonstrated that relativistic kinematics of AFM TMS permits ultra-fast energy uptake. This process is based on the relativistic DW contraction.
Recently, practical realisation of relativistic kinematics in isolated magnetic solitons has been already demonstrated \cite{Caretta2020Science}. 
In a second step,  thanks to the soliton nature of AF DWs, they can propagate along the material allowing long-range energy transport. Finally, topologically conserved collisions of two DWs which allow for annihilation processes can serve as a transfer protocol, to make use of the energy transported by the AF DWs in a fast manner. 


Notably, despite the intrinsic difficulties to measure magnetic signal in AFMs, the temperature traceability unveiled by our proposal opens the door to track experimentally the location of TMSs and their interactions in AFMs.
Our proposal also opens the door to ultra-fast energy management at the nanoscale for future nanoelectronics. AFM TMS have the potential to become  a new niche for energy transport devices based on electron's spin rather than on its charge. 

\begin{acknowledgments}
 The work of R.M.O. was partially supported by the STSM Grants from the COST Action CA17123 Ultrafast opto-magneto-electronics for non-dissipative information technology". R.M.O. would like to thank Andrew Ramsey for useful discussions. U. A. and R.R.-E. acknowledge support from the Deutsche Forschungsgemeinschaft through SFB/TRR 227  ``Ultrafast Spin Dynamics", Project A08. G.T. acknowledges the Grant-in-Aid for Scientific Research (B) (No. 17H02929) from the Japan Society for the Promotion of Science. O.C-F. acknowledges the financial support from Spanish Ministry of Science and Innovation under the grant  PID2019-108075RB-C31I00/AEI/10.13039/501100011033.
\end{acknowledgments}

\bibliography{Biblio}

\clearpage

\widetext
\clearpage
\begin{center}
\textbf{\large Supplemental Material for "Topological energy release from collision of relativistic topological solitons"}
\end{center}

\section{Atomistic Spin Model}

We perform atomistic spin dynamics simulations for the full Mn$_{2}$Au crystal structure. A unit cell is replicated along the $x$-direction 6000 times representing circa 2 $\mu$m of physical spin space. The system has periodic boundary conditions imposed along $y$ direction while open boundaries are considered along $x$ and $z$ directions. The time evolution of a unit vector spin at site $i$, $\textbf{S}_{i}$, is simulated by solving the Landau-Lifshitz-Gilbert equation:
\begin{equation}
\label{eq:LLG}
\frac{d\textbf{S}_{i}}{dt} = -\gamma\,\textbf{S}_{i}\times\textbf{H}_{i}^{\text{eff}}-
\gamma\alpha_\mathrm{G}\,\textbf{S}_{i}\times\left(\textbf{S}_{i}\times\textbf{H}_{i}^{\text{eff}}\right), 
\end{equation} 
where $\gamma$ is the gyromagnetic ratio of a free electron (2.21$\times\text{10}^{5}$m/As), $\alpha_\mathrm{G}$ is the Gilbert damping set here to 0.001 and $\textbf{H}_{i}^{\text{eff}}$ is the effective field resulting from all of the interaction energies. The configuration energy is constituted by three exchange interactions (two antiferromagnetic and one ferromagnetic), magneto crystalline energy contributions and the SO field. The total energy, $E$ is:
\begin{eqnarray}
\label{eq:Energy}
E = -2 \sum_{\langle i<j\rangle}{J_{ij}\textbf{S}_{i}\cdot\textbf{S}_{j}}-
K_{2\perp}\sum_{i}{\left(\textbf{S}_{i}\cdot\hat{\textbf{z}}\right)^{2}}- \nonumber
K_{2\parallel}\sum_{i}{\left(\textbf{S}_{i}\cdot\hat{\textbf{y}}\right)^{2}}- \\
\frac{K_{4\perp}}{2}\sum_{i}{\left(\textbf{S}_{i}\cdot\hat{\textbf{z}}\right)^{4}}-
\frac{K_{4\parallel}}{2}\sum_{i}{\left[\left(\textbf{S}_{i}\cdot\hat{\textbf{u}}_{1}\right)^{4}+
\left(\textbf{S}_{i}\cdot\hat{\textbf{u}}_{2}\right)^{4}\right]}-
\mu_{0}\mu_{S}\sum_{i}{\textbf{S}_{i}\cdot\textbf{H}_{i}^{\text{eff}}}.
\end{eqnarray}
The first term on the right-hand side is the exchange energy where $J_{ij}$ is the exchange coefficient along the considered bonds \cite{RoyPRB2016,Otxoa2020CommPhys}. The second and third terms are the uniaxial hard and easy anisotropies of strengths $K_{2\perp}$ and $K_{2\parallel}$, respectively, while the fourth and fifth terms collectively describes tetragonal anisotropy. For the in-plane part of the tetragonal anisotropy, $\textbf{u}_{1}$=$\left[110\right]$ and $\textbf{u}_{2}$=$\left[1\bar{1}0\right]$. Finally, $\mu_0$ and $\mu_{\text{s}}$ are the magnetic permeability in vacuum and the magnetic moment, respectively. We have used $\mu_{\text{s}}= 4\mu_{B}$ \cite{barthem2013revealing}, with $\mu_{B}$ being the Bohr magneton. The effective field is evaluated at each spin site in time using Eq. (\ref{eq:Energy}) as $\textbf{H}^{\text{eff}}_{i}=\frac{-1}{\mu_{0}\mu_{s}}\frac{\delta E}{\delta \textbf{S}_{i}}$. The system of equations, Eq. \eqref{eq:LLG} are solved by a fifth order Runge-Kutta Method. Material constants used are summarised in the following table:

\label{table:1}
\begin{table}[h!]
\begin{center}
	\begin{tabular}{||c c c c c c c||} 
		\hline
		$J_{i1}k_{\text{B}}^{-1}$[K] & $J_{i2}k_{\text{B}}^{-1}$[K] & $J_{i3}k_{\text{B}}^{-1}$[K] & $K_{2\perp}$[J] & $K_{2\parallel}$[J] & $K_{4\perp}$[J] & $K_{4\parallel}$[J] \\ [0.5ex] 
		\hline\hline
		-396 & -532 & 115 & -1.303$\times\text{10}^{\text{-22}}$ & 7$K_{4\parallel}$ & 2$K_{4\parallel}$ & 1.855$\times\text{10}^{\text{-25}}$ \\ 
		\hline
	\end{tabular}
\end{center}

\caption{Literature values for material parameters relevant for modelling the spin dynamics \cite{barthem2013revealing,PhysRevB.81.212409}. $k_{\text{B}}$ is Boltzmann's constant.}
\label{table:1}
\end{table}

\section{Dynamics of the self-energy of two domain walls}

In the ferromagnetic (FM) basal planes of the antiferromagnet (AFM) Mn$_2$Au, it is possible to stabilize $180^{\circ}$ Neel-like domain walls (DWs) due to the presence, mainly, of the uniaxial hard-axis anisotropy along the $z$-{\it th} spatial direction, encoded through the $ K_{2 \perp}$ parameter, which constrains the magnetization in the $ xy $ basal plane, and due to the action of the uniaxial easy-axis anisotropy along the $y$-{\it th} spatial direction, which is represented by $ K_{2 \parallel}$ (See previous section for more information). To describe one-dimensional (1D) magnetic DWs, it is convenient to introduce the Walker-like symmetric rigid profile, given by 
\begin{equation}
\phi_i \left( x, t \right) = 2 \arctan \mathrm{exp} \left[ \frac{Q_i \left( x - X_i \left( t \right) \right)}{\Delta} \right],
\label{eq:0}
\end{equation}
expression which is characterized by the DW topological charge, $Q_i$, the central position of the soliton, $ X_i $, and its DW width, $ \Delta $ \cite{schryer1974motion, rajaraman1982solitons}. We consider a system consisting of two DWs, denoted through $i=1,2$ indices, with DW topological charges and central positions $Q_{1,2}$ and $X_{1,2}$, respectively. In this case, the combined profile of the system is given by
\begin{equation}
\phi \left( x, t \right) = \sum_i \phi_i \left( x,t \right) = 2 \arctan \mathrm{exp} \left[ \frac{Q_1 \left( x - X_1 \left( t \right) \right)}{\Delta} \right]+2 \arctan \mathrm{exp} \left[ \frac{Q_2 \left( x - X_2 \left( t \right) \right)}{\Delta} \right].
\label{eq:1}
\end{equation}

The total exchange energy, $E_{\mathrm{exc}}$, composed of both magnetic textures for the case of a 1D spin chain, can be expressed as
\begin{equation}
E_{\mathrm{exc}}=a^2_0 A \, \int^{+ \infty}_{-\infty} {\left( \sum_i \partial_x \phi_i \left( x,t \right) \right)}^2 \diff x,
\label{eq:2}
\end{equation}
where $ A $ represents the effective exchange stiffness of the system, $a_0$ is the in-plane lattice constant \cite{otxoa2020walker}, and where $\partial_x$ expresses the spatial derivative along the $x$-{\it th} spatial direction of the profile of each DW, the latter being given by
\begin{equation}
 \partial_x \phi_i \left( x, t \right)= \frac{Q_i}{\Delta} \sech \left( \frac{x - X_i \left( t \right)}{\Delta} \right).
\label{eq:3}
\end{equation}

Taking into account Eq. \eqref{eq:3}, it is possible to explicitly write the terms inside the integral of Eq. \eqref{eq:2} as follows
\begin{gather}
{\left( \sum_i \partial_x \phi_i \left( x, t \right) \right)}^2=\frac{1}{\Delta^2} \left[ \sech^2  \left( \frac{ x - X_1 \left( t \right)}{\Delta} \right) + \sech^2  \left( \frac{x - X_2 \left( t \right)}{\Delta} \right) + \right. \nonumber \\ \left. 2 Q_1 Q_2 \sech \left( \frac{ x - X_1 \left(t \right) }{\Delta} \right) \sech  \left( \frac{  x - X_2 \left( t \right) }{\Delta} \right)  \right].
\label{eq:4}
\end{gather}

If this is substituted in Eq. \eqref{eq:2}, it is possible to identify different exchange-based contributions to the system.
Each DW self-energy, $ E^i_{\mathrm{exc}}$ is therefore given by
\begin{equation}
E^i_{\mathrm{exc}}=\frac{a^2_0 A}{\Delta^2} \int^{+ \infty}_{-\infty} \sech^2 \left( \frac{ x- X_i \left( t \right) }{\Delta} \right) \diff x.
\label{eq:5}
\end{equation}
An additional energy contribution comes from the exchange interaction energy between both DWs, $E^{1,2}_{\mathrm{exc}}$, which is expressed as
\begin{equation}
E^{1,2}_{\mathrm{exc}}=\frac{2Q_1Q_2 a^2_0 A}{\Delta^2} \int^{+ \infty}_{-\infty} \sech \left( \frac{ x- X_1 \left( t \right) }{\Delta} \right) \sech \left( \frac{x- X_2 \left( t \right)}{\Delta} \right) \diff x,
\label{eq:6}
\end{equation}
which allows for writing the total energy, $E_{\mathrm{exc}}$, as the sum of the following independent exchange contributions: $E_{\mathrm{exc}}=E^1_{\mathrm{exc}}+E^2_{\mathrm{exc}}+E^{1,2}_{\mathrm{exc}}$. In \nameref{section:SNII}, we focus on the interaction term between DWs (third term on the right-hand side in Eq. \eqref{eq:4}) and the resulting dynamics when subject to a spin-orbit (SO) field. 

\begin{figure*}[ht!]
\centering
\includegraphics[width=9cm]{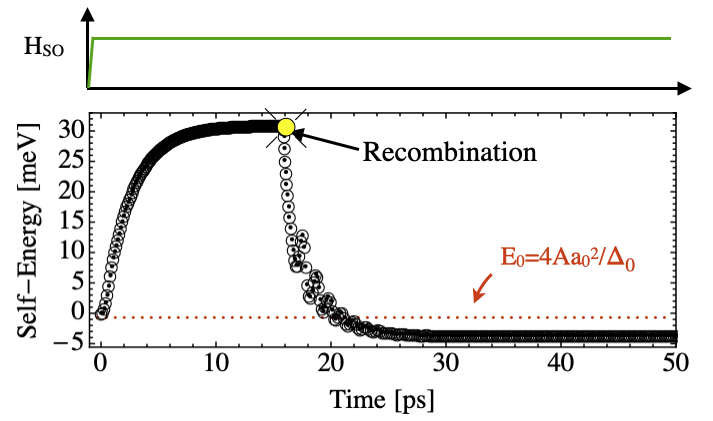}
\caption{Time dependence of the self-energy of the two antiferromagnetic domain walls (DWs) when subject to a spin-orbit field, $H_\text{SO}$, which ramps up from $H_{\mathrm{SO}}=0$ to $H_{\mathrm{SO}}=60$ mT in $10$ fs, in the case of two solitons with opposite topological charges, which entails their annihilation when they are conducted towards each other, which implies the liberation of the self-energy of both textures. The red dashed line represent the sum of the static self-energy of the two DWs, $E_0=2E^i_{\mathrm{exc}} \left( H_{\mathrm{SO}}=0 \right)=4 a^2_0 A/\Delta_0$, where $A$ represents the effective exchange stiffness of the medium, $a_0$ is the in-plane lattice constant, and where $\Delta_0$ expresses the DW width at rest, according to Eq. \eqref{eq:7}, which is taken as a zero-reference for the time evolution of the exchange energy of the system.}
\label{imS:1}
\end{figure*}

Analyzing the individual contributions to the exchange energy by the magnetic textures, expressed in Eq. \eqref{eq:5}, it is possible to obtain that
\begin{equation}
E^i_{\mathrm{exc}}=\frac{2 a^2_0 A}{\Delta},
\label{eq:7}
\end{equation}
so if the self-energy of both textures is taken into account, we can see that, regardless of their topological charge $Q_i$, both contributions will be equal, giving rise to $E^1_{\mathrm{exc}}+E^2_{\mathrm{exc}}=2 E^i_{\mathrm{exc}}=4 a^2_0 A / \Delta$, since the involved hyperbolic function is an even function. The time evolution of the self-energy, $E^i_{\mathrm{exc}}$, as a function of the relativistic contraction when subject to a spin-orbit (SO) field, $H_\mathrm{SO}$, is shown in Suppl. Fig. \ref{imS:1}.

\section{Calculation of the interaction energy between two domain walls}\label{section:SNII}

The integral given by Eq. \eqref{eq:6}, which represents the DW interaction, is more complicated than the other two terms. To solve it, we will first use the following trigonometric relationship
\begin{equation}
\cosh \left( \xi_1 \right) \cosh  \left( \xi_2 \right) = a+a \cosh \left( 2 \xi_1 \right)+b \sinh \left( 2 \xi_1 \right),
\label{eq:8}
\end{equation}
where $\xi_i= \left( x-X_i \left( t \right) \right) / \Delta$, and 
\begin{gather}
a=\frac{1}{2} \cosh \left( \frac{X_1 \left( t \right)-X_2 \left( t \right)}{\Delta} \right),  \label{eq:9} \\
b=\frac{1}{2} \sinh \left( \frac{X_1 \left( t \right)-X_2 \left( t \right)}{\Delta} \right).  \label{eq:10}
\end{gather}

The integral in Eq. \eqref{eq:6} reduces to
\begin{equation}\
E^{1,2}_{\mathrm{exc}}=\frac{2Q_1Q_2 a^2_0 A}{\Delta^2} \int^{+ \infty}_{- \infty} \frac{\diff x}{a+a \cosh \left( 2 \xi_1 \right)+b \sinh \left( 2 \xi_1 \right)}=\frac{2Q_1Q_2 a^2_0 A}{\Delta^2} \, I,
\label{eq:11}
\end{equation}
where the integral $ I $ can be rewritten as the following sum of terms
\begin{gather}
I =\frac{\Delta}{2}\int^{+\infty}_0  \left[ \frac{\diff \xi}{a+a \cosh \left( \xi \right)+b \sinh \left( \xi \right)}+ \frac{\diff \xi}{a+a \cosh \left( \xi \right)-b \sinh \left( \xi \right)} \right] \nonumber \\ 
= \frac{\Delta}{2} \left[ I \left(a, a, b \right)+ I \left( a, a, -b \right) \right]. \label{eq:12}
\end{gather}

This integral has a tabulated solution \cite{gradshteyn2014integrals}, given by
\begin{equation}
I \left(a, a, \pm b \right) = \pm \frac{1}{b} \, \mathrm{ln} \left(  \frac{a \pm b}{a} \right).
\label{eq:13}
\end{equation}

Thus, using the relation $2 \, \mathrm{atanh} \left( x \right)= \text{ln} \left[ \left( 1+x \right) / \left( 1-x \right) \right]$, the integral can be rewritten as
\begin{equation}
I=2 \left( X_1 \left( t \right) - X_2 \left( t \right) \right) \csch \left( \frac{X_1 \left( t \right)-X_2 \left( t \right)}{\Delta} \right),
\label{eq:14}
\end{equation}
and, subsequently, Eq. \eqref{eq:11} can be expressed as
\begin{equation}
E^{1,2}_{\mathrm{exc}}=\frac{4 Q_1Q_2 a^2_0 A \left( X_1 \left( t \right) - X_2 \left( t \right)\right)}{\Delta^2} \csch \left( \frac{X_1 \left( t \right)-X_2 \left( t \right)}{\Delta} \right).
\label{eq:15}
\end{equation}

Summarizing all the exchange contributions of the system, taking into account Eqs. \eqref{eq:7} and \eqref{eq:15}, Eq. \eqref{eq:2} takes the functional form
\begin{equation}
E_{\mathrm{exc}}=\frac{4 a^2_0 A}{\Delta}+\frac{4 Q_1Q_2 a^2_0 A \left( X_1 \left( t \right) - X_2 \left( t \right)\right)}{\Delta^2} \csch \left( \frac{X_1 \left( t \right)-X_2 \left( t \right)}{\Delta} \right).
\label{eq:16}
\end{equation}

\begin{figure*}[!ht]
\centering
\includegraphics[width=13cm]{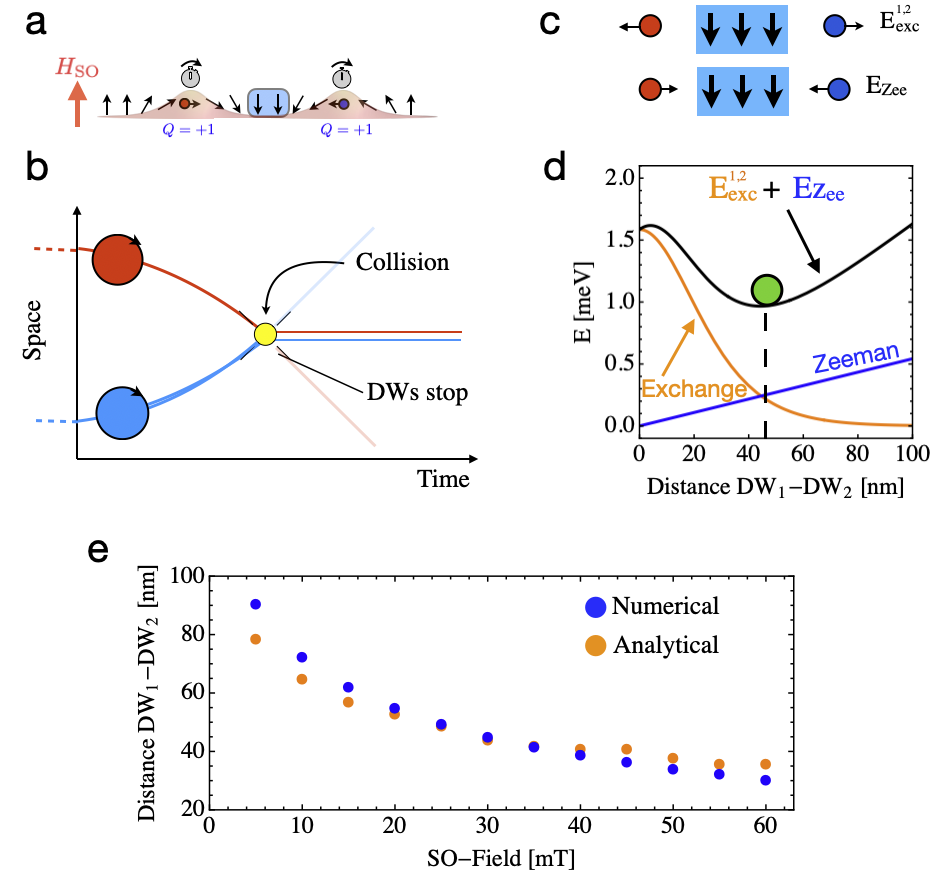}
\caption{a) Sketch of two domain walls (DWs) with the same chirality $Q=+1$ under the action of a spin-orbit (SO) field, $H_{\mathrm{SO}}$. They move driven by SO-field antiparallel to the central magnetic domain (blue box). b) DWs move until collision.  After the collision, both DWs remain at an equilibrium distance from each other. c) Schematic illustration of the role played by the exchange, $E^{1,2}_{\mathrm{exc}}$, and the Zeeman, $E_{\mathrm{Zee}}$, energies. While the exchange energy tries to separate DW$_1$ and DW$_2$, the Zeeman energy forces them to stay as close as possible. d) Zeeman (blue line) and exchange (orange line) energies as a function of the relative distance between DW$_1$ and DW$_2$ for an applied SO-field of 20 mT. e) Comparison between analytical Eq. \eqref{eq:17} and numerically extracted stable distances between DW$_1$ and DW$_2$ as a function of the applied SO-field}.
\label{imS:2}
\end{figure*}

As we have shown in the main text, when both DWs have opposite topological charge ($Q_1+Q_2=0$), there is a continuous transformation that permits both DWs to recombine giving rise to a uniform magnetic state. However, when $Q_1+Q_2=2$ (See Suppl. Fig. \ref{imS:2}a and b), there exists a region between both DWs whose polarisation is antiparallel to $H_\mathrm{SO}$. As a consequence, the Zeeman energy, $E_\text{Zee}$, pushes the DWs together in order to minimise the related energy (See Suppl. Fig. \ref{imS:2}c). On the other hand, there also exists a repulsive interaction between DW$_1$ and DW$_2$ due to the exchange interaction, $E^{1,2}_\mathrm{exc}$, given by Eq. \eqref{eq:15}. This results into an expansion of the magnetic domain separating both DWs. Note that the central spins of each DW are antiparallel to each other, which implies that, to minimize the exchange energy, both magnetic textures have to be as far as possible from the other. The competition between these two forces, which leads to a stable distance among the two DWs, as shown in Suppl. Fig. \ref{imS:2}b, can be expressed as
\begin{equation}
\Delta E=\mu_0 M_{\mathrm{s}} a^2_0 H_{\mathrm{SO}} \left( X_1-X_2 \right)+\frac{4 Q_1Q_2 a^2_0 A \left( X_1 - X_2 \right)}{\Delta^2} \csch \left( \frac{X_1 -X_2}{\Delta} \right),
\label{eq:17}
\end{equation}
where $ \mu_0 $ is the vacuum permeability and $ M_{\mathrm{s}} $ represents the volumetric saturation magnetization \cite{otxoa2020walker}. Suppl. Fig. \ref{imS:2}d, shows the dependence of the Zeeman and exchange energies as a function of the relative distance between two DWs (DW$_1$ and DW$_2$) with same topological charge for an applied SO-field of 20 mT. As the SO-field and magnetic domain between the two DWs is antiparallel, the minimum energy corresponds to a zero distance between the DWs. The opposite happens with the exchange energy, the smaller is the distance between the two DWs, the larger is the exchange energy ($A>0$). One can see (black dashed line in Suppl. Fig. \ref{imS:2}d) that when accounting for the exchange and the Zeeman energies, it appears a global minimum which corresponds to a stable configuration distance between DW$_1$ and DW$_2$. A comparison between the stable distance among the two DWs as a function of the applied SO-field extracted from numerical simulations and Eq. \eqref{eq:17} is shown in Suppl. Fig. \ref{imS:2}e. It can be observed that starting for the larger SO-field (60 mT), the stable distance increases as the SO-field is reduced, meaning that the Zeeman force needs more extension of the magnetic domain to compensate the repulsion between the DWs.

\section{Equilibrium distance between domain walls and relaxation time of breather mode}

\begin{figure*}[!ht]
\centering
\includegraphics[width=8cm]{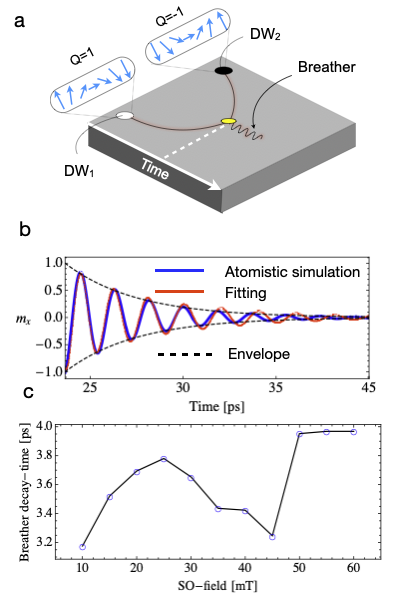}
\caption{a) Schematic illustration of two domain walls (DW$_1$ and DW$_2$) with opposite topological charge ($Q_1=-Q_2$) colliding at certain time instant, $t$, forming a bound state known as breather. b) Time evolution of the $m_x$ component at a fixed point $x_0$ of annihilation, extracted along the $(x_0,t)$. The oscillations are well fitted by Eq. \eqref{eq:29}. c) Breather decay-time dependence with the spin-orbit field}.
\label{im:5}
\end{figure*}

The topological charge is an invariant of the system irrespective of the number of magnetic textures. As we show in Suppl. Fig. \ref{im:5}a, when two domain DWs (DW$_1$ and DW$_2$) with opposite winding number collide, they recombine because the overall topological charge, $Q=Q_1+Q_2$ is zero. From the energetic point of view, each DW can be interpreted as an effective potential for the other one. When the velocity at which the collision occurs is not sufficiently large, both DWs become trapped inside the other DW effective potential leading to an oscillatory and localized spin excitation usually called breather \cite{rajaraman1982solitons}. The dampening process as well as the frequency of the oscillation can be mapped into a damped harmonic oscillator reproducing nicely the convoluted oscillations obtained from numerical simulations, see Suppl. Fig. \ref{im:5}b. For this purpose, it is necessary to introduce the functional form of the damped harmonic oscillator, which is given by
\begin{equation}
m_x=B \cos \left( \omega t-\varphi \right) e^{-dt},
\label{eq:29}
\end{equation}
where $B$ represents the amplitude of the oscillation (being the $x$-{\it th} magnetization component, $m_x$, bounded between $[-1,1]$), $\omega$ represents the oscillation frequency, $\varphi$ is the phase, and $d$ corresponds to the characteristic decay-time for the breather.

Suppl. Fig. \ref{im:5}c illustrates the dependence of the decay time, $b$, as a function of the SO-field. It can be deduced that irrespective of the velocity at which the collision takes place, the time devoted by the spin system to release the energy stored in each topological magnetic soliton is only dependent of the intrinsic parameters of the system such as the damping, saturation magnetisation and configurational energy terms

\end{document}